\documentclass[conference]{IEEEtran}

\ifCLASSINFOpdf
\usepackage[pdftex]{graphicx}

\else

\fi

\usepackage[cmex10]{amsmath}
\usepackage{amssymb}   
\usepackage{amsxtra}
\usepackage{amscd}
\usepackage{amsthm}
\usepackage{textcomp}
\usepackage{graphicx}

\usepackage{balance}

\usepackage{array}  

\usepackage{multirow}  

\usepackage{amsmath}
\usepackage{cite}

\newcommand\blfootnote[1]{%
  \begingroup
  \renewcommand\thefootnote{}\footnote{#1}%
  \addtocounter{footnote}{-1}%
  \endgroup
}

\usepackage{url}

\usepackage{color,soul} 
\soulregister\cite7
\soulregister\ref7
\soulregister\pageref7

\makeatletter

\def\ps@IEEEtitlepagestyle{%
	\def\@oddfoot{\mycopyrightnotice}%
	\def\@evenfoot{}%
}
\def\mycopyrightnotice{%
	{\footnotesize  \hspace*{5cm}2019 European Conference on Networks and Communications (EuCNC)\hfill}
	\gdef\mycopyrightnotice{}
}

\begin{document}
%

\title{{\huge Transmission Through Large Intelligent Surfaces: \\A New Frontier in Wireless Communications}}

\author{Ertugrul Basar}

\author{\IEEEauthorblockN{Ertugrul Basar}
	\IEEEauthorblockA{Communications Research and Innovation Laboratory (CoreLab), Department of Electrical and Electronics Engineering\\
		Ko\c{c} University, Sariyer 34450, Istanbul, Turkey. E-mail: ebasar@ku.edu.tr}\vspace*{-0.62cm}}

\maketitle

\begin{abstract}
In this paper, transmission through large intelligent surfaces (LIS) that intentionally modify the phases of incident waves to improve the signal quality at the receiver, is put forward as a promising candidate for future wireless communication systems and standards. For the considered LIS-assisted system, a general mathematical framework is presented for the calculation of symbol error probability (SEP) by deriving the distribution of the received signal-to-noise ratio (SNR). Next, the new concept of using the LIS itself as an access point (AP) is proposed. Extensive computer simulation results are provided to assess the potential of LIS-based transmission, in which the LIS acts either as an intelligent reflector or an AP with or without the knowledge of channel phases. Our findings reveal that LIS-based communications can become a game-changing paradigm for future wireless systems.  \blfootnote{This work was supported in part by the Scientific and Technological Research Council of Turkey (TUBITAK) under Grant 117E869, the Turkish Academy of Sciences (TUBA) GEBIP Programme, and the Science Academy BAGEP Programme. Codes available at \url{https://corelab.ku.edu.tr/tools}.}
  
\end{abstract}

%
\IEEEpeerreviewmaketitle


\section{Introduction}

The first commercial fifth generation (5G) wireless networks have been already deployed in certain countries while the first 5G compatible handsets are expected to be available during 2019. Although the initial stand-alone 5G standard, which brings more flexibility into the system design by exploiting millimeter-waves and multiple orthogonal frequency division multiplexing numerologies, has been completed during 2018, researchers are relentlessly exploring the potential of  emerging technologies for later releases of 5G. These potential technologies include non-orthogonal multiple access, optical wireless communications and hybrid optical/radio frequency (RF) solutions, alternative waveforms, low-cost massive multiple-input multiple-output (MIMO) systems, terahertz communications, and new antenna technologies. Even though future 6G technologies look like as the extension of their 5G counterparts at this time \cite{6G}, new user requirements, completely new applications/use-cases, and new networking trends of 2030 and beyond may bring more challenging communication engineering problems, which necessitate radically new communication paradigms in the physical layer.

Within this context, there has been a growing interest in controlling the propagation environment in order to increase the quality of service for wireless communications. Schemes such as media-based modulation \cite{Khandani_conf1,MBM_TVT,Basar_2019}, spatial scattering modulation \cite{SSM}, and beam index modulation (IM) \cite{BIM}, which belong to the vast IM family \cite{Basar_2017}, use the variations in the signatures of received signals by exploiting reconfigurable antennas or scatterers to transmit additional information bits in rich scattering environments. On the other hand, large intelligent surfaces/walls/reflect-arrays/metasurfaces are smart devices that control the propagation environment with the aim of improving the coverage and signal quality \cite{Akyildiz_2018}.

It is worth noting that the \textit{large intelligent surface (LIS)}-based transmission concept is completely different from existing MIMO, beamforming, amplify-and-forward relaying, and backscatter communication paradigms, where the large number of small, low-cost, and passive elements on a LIS only reflect the incident signal with an adjustable phase shift without requiring a dedicated energy source for RF processing, decoding, encoding, or retransmission. Inspired by the definition of software-defined radio, which is given as ``radio in which some or all of the physical layer functions are software defined" and considering the interaction of the intelligent surface with incoming waves in a software-defined fashion, we may also use the term of \textit{software-defined surface (SDS)} for these intelligent surfaces.

The concept of intelligent walls is proposed in one of the early works by utilizing active frequency selective surfaces to control the signal coverage \cite{Subrt_2012}. Alternative to beamforming techniques that require large number of antennas to focus the transmitted or received signals, the concept of smart reflect-arrays is proposed in \cite{Tan_2016}. It has been also demonstrated that reflect-arrays can be used  effectively to change the phase of reflected signals without buffering or processing the incoming signals and the received signal quality can be enhanced by adjusting the phase shift of each element on the reflect-array. As an evolution of massive MIMO systems, the LIS concept is proposed in \cite{Hu_2018} by exploiting the whole contiguous surface for transmitting and receiving. The authors of \cite{Huang_2018,Huang_2018_2,Huang_2019} focused on a downlink transmission scenario through a LIS to support multiple users and investigated sum-rate and energy efficiency maximization problems. Low complexity algorithms are also considered for the encountered non-convex optimization problems to obtain the optimum reflector phases. Recently, a joint active and passive beamforming problem is investigated in \cite{Wu_2018} and \cite{Wu_2018_2}, and the user's average received power is investigated.

Against this background, this paper first provides a mathematical framework for the error performance analysis of LIS-based communication systems. For the first time in the literature, we investigate the effect of number of reflecting elements, modulation orders, and blind phases on the error performance and provide interesting asymptotic results depending on different signal-to-noise ratio (SNR) regimes. Second, inspired by the promising potential of LIS-based transmission, we propose the concept of using the LIS itself as an access point (AP) by exploiting an unmodulated carrier that is generated by a nearby RF signal generator and transmitted towards the LIS. In this scheme, reflector phases are used not only for SNR maximization but also for information transmission. It has been shown by extensive computer simulations as well as theoretical derivations that a LIS can be used effectively both as a reflector and as an AP in future 6G wireless networks.

The rest of the paper is organized as follows. In Section II, we introduce the system model of the LIS-based communication scheme and evaluate its symbol error probability (SEP). Section III introduces the new LIS-based AP concept. Computer simulation results and comparisons are given in Section IV. Finally, conclusions are given in Section V.

\section{Transmission Through LIS: System Model \& Error Performance Analysis}

\begin{figure}[!t]
	\begin{center}
	\includegraphics[width=1\columnwidth]{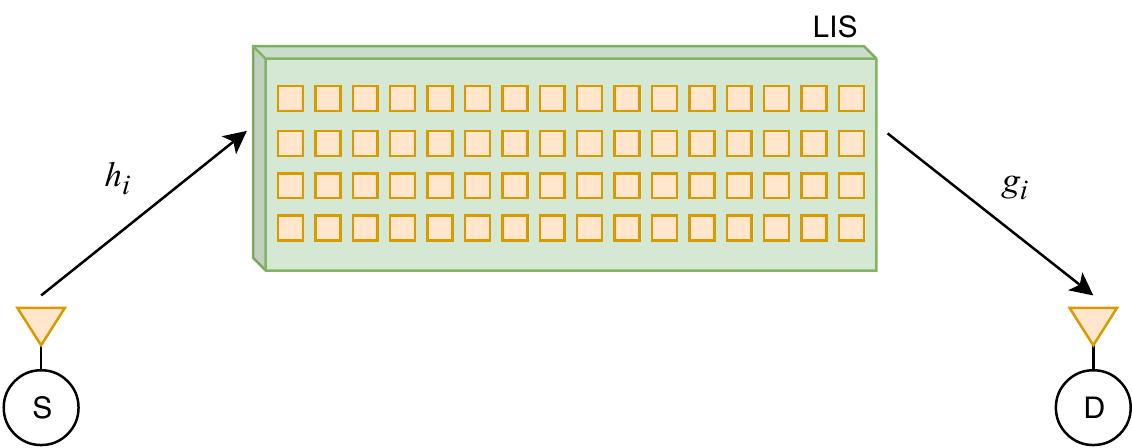}
		\vspace*{-0.6cm}\caption{Transmission through a LIS in a dual-hop communication scenario without a line-of-sight path between S and D.}\vspace*{-0.3cm}
	\end{center}
\end{figure}

In this section, we present the system model of the generic LIS-based scheme and provide a unified framework for the calculation of its theoretical SEP. The block diagram of the considered LIS-based transmission scheme is shown in Fig. 1, where $h_i$ and $g_i$ respectively represent the fading channel between the single-antenna source (S) and the LIS, and the LIS and the single-antenna destination (D). Under the assumption of Rayleigh fading channels, we have $h_i,g_i \sim \mathcal{CN}(0,1)$, where $\mathcal{CN}(0,\sigma^2)$ stands for the complex Gaussian distribution with zero mean and $\sigma^2$ variance. We assume that the LIS is in the form of a reflect-array comprising $N$ simple and reconfigurable reflector elements, and controlled by a communication-oriented software. We investigate two different implementation scenarios considering the knowledge of channel phases at the LIS: i) intelligent transmission and ii) blind transmission.

\subsection{Intelligent Transmission Through LIS}

\begin{figure}[!t]
	\begin{center}
		\includegraphics[width=0.83\columnwidth]{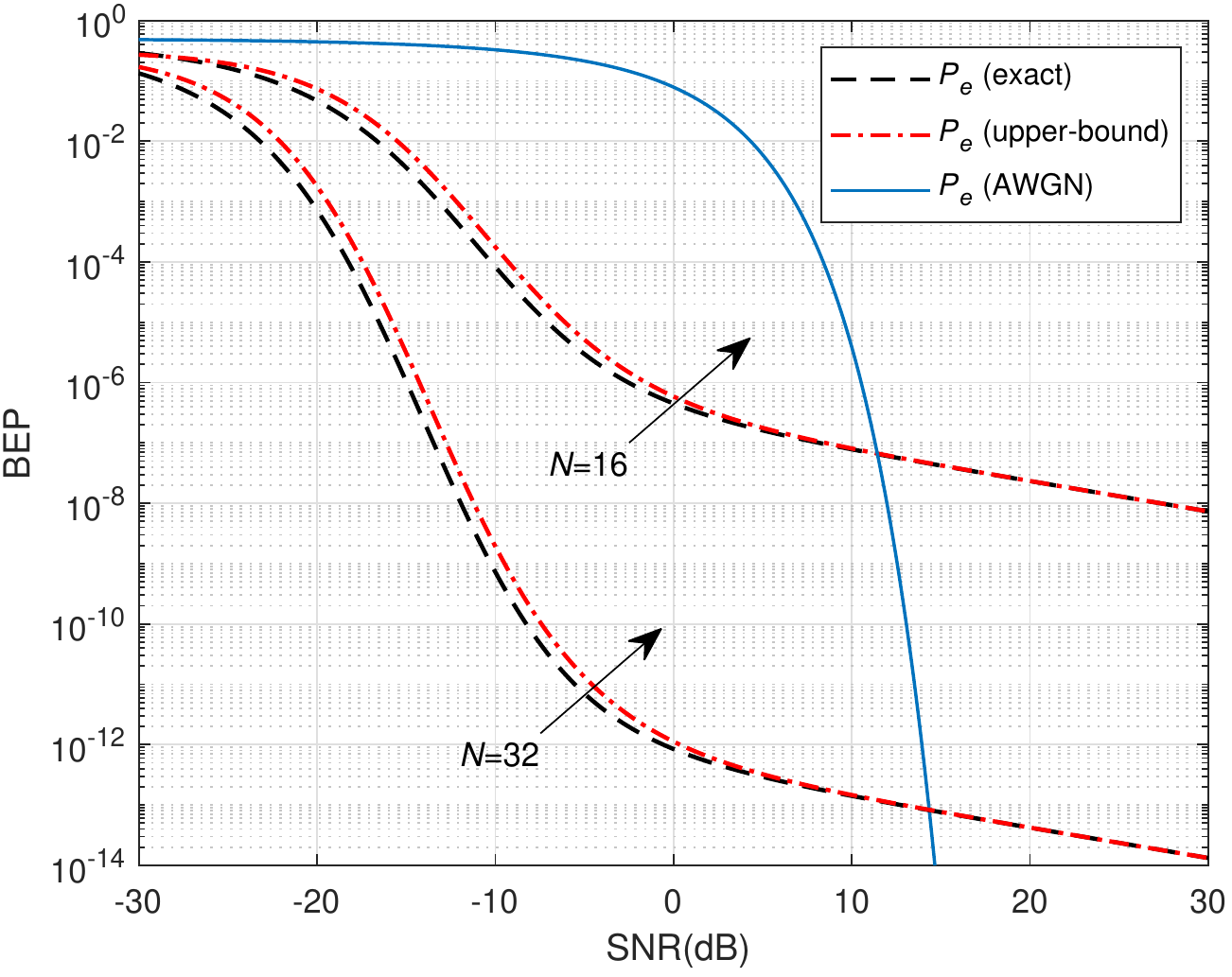}
		\vspace*{-0.35cm}\caption{Theoretical average BEP of the LIS-based scheme for $N=16$ and $N=32$ with BPSK.}\vspace*{-0.4cm}
		\label{N_16_32_BER}
	\end{center}
\end{figure}

For the case of slowly varying and flat fading channels, the received baseband signal reflected through the LIS with $N$ passive elements can be expressed as 
\begin{equation}\label{eq:1}
r=\left[ \sum_{i=1}^{N} h_i e^{j \phi_i} g_i \right] x + n 
\end{equation}
where $\phi_i$ is the adjustable phase induced by the $i$th reflector of the LIS, $x$ stands for the data symbol selected from $M$-ary phase shift keying/quadrature amplitude modulation (PSK/QAM) constellations and $n \sim \mathcal{CN}(0,N_0)$ is the additive white Gaussian noise (AWGN) term. Here, we have $h_i = \alpha_i e^{-j \theta_i}$ and $g_i = \beta_i e^{-j \psi_i}$ in terms of channel amplitudes and phases. From \eqref{eq:1}, the instantaneous SNR at D is calculated as
\begin{equation}
\gamma =  \frac{\left|  \sum_{i=1}^{N} \alpha_i \beta_i e^{j (\phi_i-\theta_i - \psi_i   ) } \right|^2 E_s }{N_0} 
\end{equation}
where $E_s$ is the average transmitted energy per symbol. It is easy to show that $\gamma$ is maximized by eliminating the channel phases with the help of the LIS as $\phi_i = \theta_i + \psi_i$ for $i=1,\ldots,N$, which requires the knowledge of channel phases at the LIS. This is verified by the identity $\left| \sum_{i=1}^{N} z_i e^{j \xi_i}\right| ^2 = \sum_{i=1}^{N} z_i^2 + 2 \sum_{i=1}^{N} \sum _{k=i+1}^N  z_i z_k \cos(\xi_i - \xi_k) $, which can be maximized by ensuring $\xi_i =\xi$ for all $i$. With the help of the LIS through intelligent reflection of the incoming electromagnetic waves, the maximized instantaneous received SNR is expressed as
\begin{equation}
\gamma =\frac{\left( \sum_{i=1}^{N} \alpha_i \beta_i   \right)^2 E_s }{N_0}= \frac{ A^2 E_s }{N_0}.
\end{equation}
Noting that $\alpha_i$ and $\beta_i$ are independently Rayleigh distributed random variables (RVs) and $\mathrm{E}[\alpha_i \beta_i]= \frac{\pi}{4}$, $\mathrm{VAR}[\alpha_i \beta_i] = 1-\frac{\pi^2}{16}$, for sufficiently large number of reflecting elements $N \gg 1$, according to the central limit theorem (CLT), $A$ follows Gaussian distribution with the following parameters: $\mathrm{E}[A]=\frac{N \pi}{4} $ and $\mathrm{VAR}[A]=N\left(1-\frac{\pi^2}{16} \right) $. Then, it is observed that $\gamma$ is a non-central chi-square RV with one degree of freedom and has the following moment generating function (MGF) \cite{Proakis}:
\begin{equation}\label{MGF_1}
M_{\gamma}(s)= \left( \dfrac{1}{1-\frac{sN(16-\pi^2)E_s}{8N_0}}\right) ^{\frac{1}{2}}  \! \!\exp\left( \dfrac{ \frac{sN^2 \pi^2 E_s}{16 N_0}}{1-\frac{sN(16-\pi^2)E_s}{8N_0}} \right). 
\end{equation}
Furthermore, the average received SNR becomes $E\left[ \gamma \right]=\frac{( N^2 \pi^2+ N(16-\pi^2)) E_s}{16N_0}$, which is proportional to $N^2$. From \eqref{MGF_1}, we can obtain the average SEP for $M$-PSK signaling as \cite{Simon}
\begin{equation}\label{5}
P_e=\frac{1}{\pi} \int_{0}^{(M-1)\pi/M} M_{\gamma} \left( \frac{-\sin^2 (\pi/M)}{\sin^2 \! \eta}\right) d\eta
\end{equation}
which simplifies to the following for binary PSK (BPSK):
\begin{equation}\label{SEP_1}
P_e= \frac{1}{\pi}  \! \int_{0}^{\pi/2} \! \! \left( \dfrac{1}{1+\frac{N(16-\pi^2)E_s}{8 \sin^2 \! \eta N_0}}\right) ^{\!\frac{1}{2}}  \! \!\! \exp\left( \dfrac{ -\frac{N^2 \pi^2 E_s}{16  \sin^2 \! \eta N_0 }}{1+\frac{N(16-\pi^2)E_s}{8 \sin^2 \! \eta N_0}} \right) \! \! d\eta. 
\end{equation}
In order to gain further insights, \eqref{SEP_1} can be upper bounded by letting $\eta=\pi/2$ as
\begin{equation}\label{SEP_1_UB}
P_e \le \frac{1}{2}   \left( \dfrac{1}{1+\frac{N(16-\pi^2)E_s}{8N_0 }}\right) ^{\!\frac{1}{2}}  \! \!\! \exp\left( \dfrac{ -\frac{N^2 \pi^2 E_s}{16 N_0 }}{1+\frac{N(16-\pi^2)E_s}{8N_0}} \right). 
\end{equation}
In Fig. \ref{N_16_32_BER}, we plot the average bit error probability (BEP) of the LIS-based scheme from \eqref{SEP_1} and \eqref{SEP_1_UB} for $N=16$ and $N=32$ with respect to $E_s/N_0$. As seen from Fig. \ref{N_16_32_BER}, the LIS-based scheme achieves significantly better BEP performance compared to the classical BPSK scheme operating over the pure AWGN channel. In other words, a LIS can convert a hostile wireless fading environment into a super communication channel that provides very low BEP at extremely low SNR values through the smart adjustment of reflector phases. The following remark explains this phenomenon.

\textit{Remark}: As seen from Fig. \ref{N_16_32_BER}, the average BEP curves have a waterfall region and a saturation region. We observe that for $\frac{N E_s}{N_0} \ll 10 $, from \eqref{SEP_1_UB}, $P_e$ becomes proportional to
\begin{equation} \label{Approx_1}
P_e \propto \exp \left( -\frac{N^2 \pi^2 E_s}{16 N_0 } \right) 
\end{equation} 
which explains the superior BEP performance of the LIS-based scheme. In this region, although the SNR $\left(E_s/N_0\right) $ is relatively low, due to the $N^2$ term in the exponent, considerably low BEP values are possible, particularly with increasing $N$. On the other hand, for  $\frac{N E_s}{N_0} \gg 1 $, \eqref{SEP_1_UB} can be approximated as
\begin{align}
P_e & \propto \left(\frac{N (16-\pi^2)E_s}{8 N_0} \right)^{-\frac{1}{2}}  \exp \left( -\frac{N \pi^2 }{2(16- \pi^2) } \right) 
\end{align}
which explains the saturated BEP performance for high SNR values due to $-\frac{1}{2}$ exponent of the SNR. However, the average BEP still decays exponentially with respect to $N$ and significant reductions are possible in $P_e$ by increasing $N$. $\hspace{1cm} \blacksquare$

In Fig. 3, we show the bit error rate (BER) performance of the LIS-based scheme for different number of reflecting elements $(N)$ and BPSK signaling. As seen from Fig. 3, our theoretical approximation in \eqref{SEP_1} using the CLT is considerably accurate for increasing $N$ values. Furthermore, we observe that doubling $N$ provides approximately $6$ dB improvement (four-fold decrease) in the required SNR at the waterfall region to achieve a target BER, which can be easily verified  from \eqref{Approx_1}.

\begin{figure}[!t]
	\begin{center}
		\includegraphics[width=0.81\columnwidth]{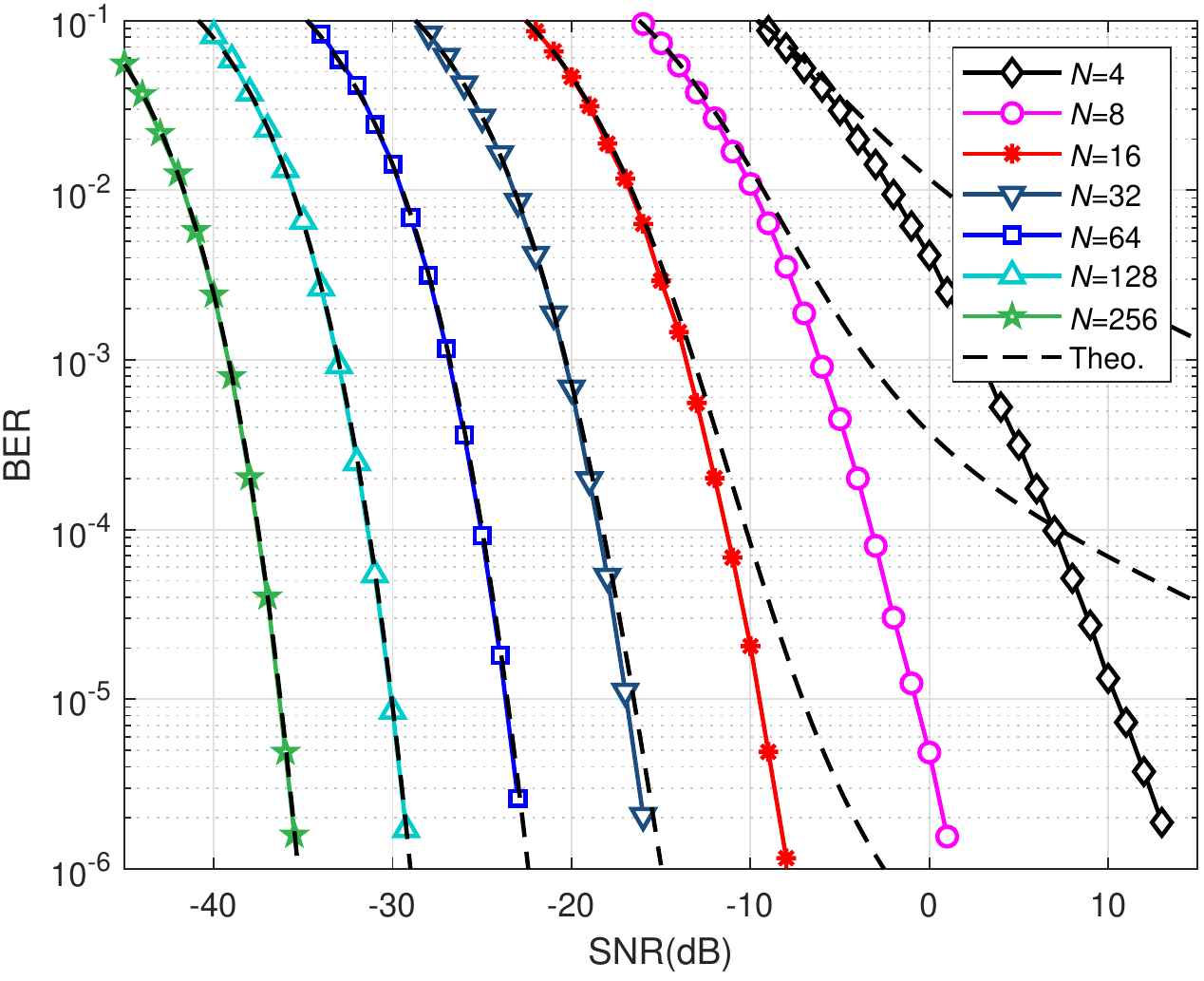}
		\vspace*{-0.35cm}\caption{Simulated BER performance of the LIS-based scheme with varying number of reflecting elements for BPSK with theoretical results of \eqref{SEP_1}.}\vspace*{-0.4cm}
		\label{Sim1}
	\end{center}
\end{figure}

Using the MGF of the received instantaneous SNR $(M_{\gamma}(s))$, we can also obtain the average SEP for square $M$-QAM constellations as \cite{Simon}
\begin{align}\label{M_QAM}
P_e = \,\, & \, \frac{4}{\pi} \left(1-\frac{1}{\sqrt{M}} \right) \int_{0}^{\pi/2} M_{\gamma} \left(  \frac{-3}{2(M-1) \sin^2 \! \eta}\right) d\eta  \nonumber \\
&\hspace*{-0.7cm}- \frac{4}{\pi} \left(1-\frac{1}{\sqrt{M}} \right)^2 \int_{0}^{\pi/4} M_{\gamma} \left(  \frac{-3}{2(M-1) \sin^2 \! \eta}\right) d \eta.  
\end{align}
Removing the integrals by letting $\eta=\pi/2$ and $\eta=\pi/4$ in the first and second terms of \eqref{M_QAM}, we can obtain a tight upper-bound on the average SEP. Under the assumption of $\frac{N E_s}{N_0} \ll 10 $ (at the SNR region of interest), the average SEP can be expressed as
\begin{equation}\label{Eq:10}
P_e \propto  \exp \left( -\frac{3N^2 \pi^2 E_s}{32(M-1) N_0}\right)  
\end{equation}
where we ignored the second exponential term coming from \eqref{M_QAM} due to its relatively larger exponent. Since $M$ appears in the exponent of  \eqref{Eq:10}, the LIS-based scheme also suffers from a degradation in error performance with increasing modulation orders although benefiting from the $N^2$ term. 

\subsection{Blind Transmission Through LIS}
In this case, the LIS given in Fig. 1 does not have the knowledge of channel phases $\theta_i$ and $\psi_i$, and consequently, cannot eliminate these phase terms to maximize the received SNR. Without loss of generality, assuming $\phi_i=0$ for $i=1,2,\ldots,N$, the received signal becomes\footnote{It is worth noting that the case of $N=1$ is equivalent to the well-known cascaded Rayleigh fading.}
\begin{equation}
r=\left[ \sum_{i=1}^{N} h_i  g_i \right] x + n =H x + n. 
\end{equation}
For this blind scheme, the CLT can be also applied for large $N $, and considering $H \sim \mathcal{C N}(0,N)$, the MGF of the received SNR is obtained as $ M_{\gamma}(s)=(1-\frac{sNE_s}{N_0})^{-1} $.  
Following the same steps above, BEP of the blind LIS-based scheme can be expressed for binary signaling as
\begin{equation}\label{eq:14}
P_e = \frac{1}{\pi} \int_{0}^{\pi/2} \! \! \left( \! \frac{1}{1 \!+ \! \frac{N E_s}{ \sin^2 \! \eta N_0}}  \! \right)  \!  \! d\eta = \frac{1}{2} \left( \!1  \!- \! \sqrt{\frac{ \frac{NE_s}{N_0}}{1  \!+  \!\frac{NE_s}{N_0}}} \,\right) 
\end{equation}
where an $N$ times SNR gain is obtained compared to point-to-point transmission over Rayleigh fading channels.

\section{The New Design: LIS As an Access Point}
Considering the promising potential of the LIS-based concept discussed in the previous section, we propose the new paradigm of transmitting information by the  LIS itself. In other words, the LIS plays the role of an AP (source) in our communication scenario, however, it is again consists of only low-cost and passive reflector elements. In this setup, the LIS can be connected to the network over a wired link or optical fiber, and can support transmission without RF processing.

The block diagram of the proposed LIS-based concept is shown in Fig. 4, where the channel between the LIS and D is modeled by $g_i = \beta_i e^{-j \psi_i}$. In this scenario, the LIS is supported by a nearby RF signal generator or contains an attachment that transmits an unmodulated carrier signal $\cos(2\pi f_c t)$ at a certain carrier frequency $f_c$ towards the LIS. Here, the unmodulated carrier can be easily generated by an RF digital-to-analog converter with an internal memory and a power amplifier \cite{Mesleh_2018}, and information bits are conveyed only through the adjustment of reflector-induced phases of the LIS. We also assume that the RF source is close enough to (or a part of/an attachment to) the  LIS and its transmission is not affected by fading. Depending on the knowledge of channel driven phase terms, this concept can be realized in two different ways: i) intelligent AP and ii) blind AP.

\begin{figure}[!t]
	\begin{center}
		\includegraphics[width=0.9\columnwidth]{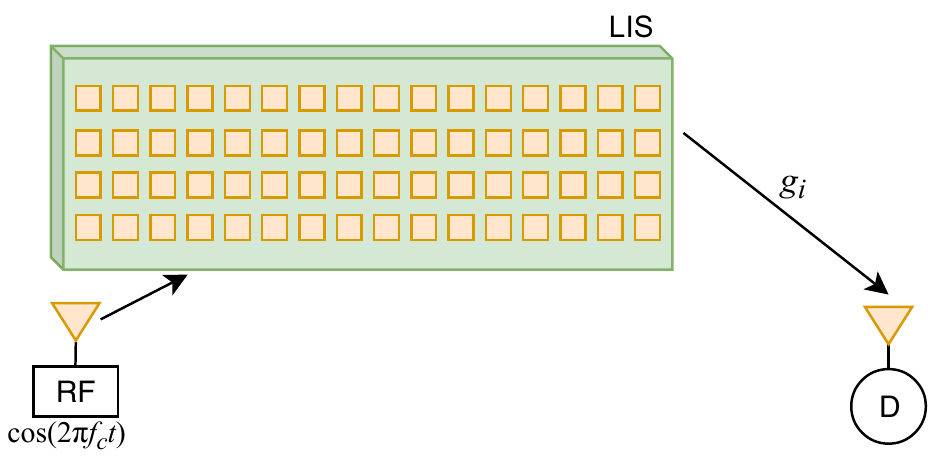}
		\vspace*{-0.35cm}\caption{The new concept: Using the LIS as an access point.}\vspace*{-0.3cm}
	\end{center}
\end{figure}

\subsection{Intelligent Access Point-LIS}
For this communication scenario, LIS-induced phases themselves carry information in addition to the intelligent reflection that improves the received SNR. In other words, the LIS adjusts the phases of its reflector elements with the aim of not only cancelling the channel phase terms to maximize the received SNR but also properly aligning the reflected signals in the 2D plane to form a virtual $M$-ary signal constellation. For this model, the received baseband signal is given as
\begin{equation}
r=\sqrt{E_s}\left[ \sum_{i=1}^N g_i e^{j \phi_i}  \right] + n 
\end{equation}
where $E_s$ is the average transmitted signal energy of the unmodulated carrier and $\phi_i$ is the reconfigurable phase induced by the $i$th reflector of the LIS. We assume that a total of $\log_2(M)$ bits are transmitted for each signaling interval by adjusting reflector phases as $\phi_i=\psi_i + w_m$, where $w_m, m\in \left\lbrace 1,2,\ldots,M \right\rbrace $ is the common additional phase term induced by the LIS to carry the information of the $m$th message. In light of this, the received signal can be expressed as
 \begin{equation}\label{signal}
 r=\sqrt{E_s}\left[ \sum_{i=1}^N \beta_i   \right] e^{j w_m} + n=\sqrt{E_s} B e^{j w_m} + n .
 \end{equation}
It is worth noting that this signal model resembles that of PSK signaling over a super-channel $B$. Consequently, to minimize the average SEP, the information phases $w_1,w_2,\ldots,w_M$ of this $M$-ary signaling scheme should be selected as in the classical $M$-PSK scheme. This can be verified by the conditional pairwise error probability (CPEP) for the transmission of message $ k $ $(w_k)$ and its erroneous detection as message $ l $ $(w_l)$, which can be calculated as follows for $k,l \in \left\lbrace 1,2,\ldots,M \right\rbrace $: 
\begin{align}
P_{e\left| \right. \! B}&=P\left( \left|r- \sqrt{E_s}B e^{j\omega_l} \right|^2 < \left|r- \sqrt{E_s} B e^{j\omega_k} \right|^2 \right) \nonumber \\
&\hspace*{-0.65cm}= P\left( \Re\left\lbrace r^* \sqrt{E_s} B (e^{j\omega_k}-e^{j\omega_l}) \right\rbrace <0  \right)  \nonumber \\
&\hspace*{-0.65cm} = P \left( E_s B^2  \! \left(1\!-\!\cos (\omega_l\!-\!\omega_k) \right)\right.  \nonumber \\
&\hspace*{-0.2cm}+\left. \Re\left\lbrace n^*\sqrt{E_s}  B (e^{j\omega_k}\!-\! e^{j\omega_l}) \right\rbrace <0   \right)  = P(D<0).
\end{align}
Here, considering the fact that $D \sim \mathcal{N}(m_D,\sigma_D^2)$, where $m_D=E_s B^2  \left(1-\cos (\omega_l-\omega_k) \right)$ and $\sigma_D^2 = N_0 E_s B^2  \left(1-\cos (\omega_l-\omega_k) \right) $, we obtain
\begin{equation}
P_{e\left| \right. \! B}= Q\left( \sqrt{\frac{E_s B^2 \left(1-\cos (\omega_l-\omega_k) \right)}{N_0}} \right) 
\end{equation}
which can be minimized with uniformly arranged phases,  that is, $w_m=2\pi(m-1)/M$ for $m=1,2,\ldots,M$.

In light of the above discussion, the instantaneous received SNR can be calculated for the model of \eqref{signal} as
\begin{equation}\label{key}
\gamma = \frac{E_s B^2}{N_0}.
\end{equation}
Considering the CLT for large $N$ and Rayleigh distribution of $\beta_i$ with mean $\sqrt{\pi}/2$ and variance $(4-\pi)/4$, we obtain $B\sim \mathcal{N}(m_B,\sigma_B^2)$, where $m_B=N\sqrt{\pi}/2$ and $\sigma_B^2=N(4-\pi)/4$. Consequently, the MGF of $\gamma$ is obtained as
\begin{equation}\label{MGF_2}
M_{\gamma}(s)= \left( \dfrac{1}{1-\frac{sN(4-\pi)E_s}{2N_0}}\right) ^{ \! \frac{1}{2}}  \! \!\exp\left( \dfrac{ \frac{sN^2 \pi E_s}{4 N_0}}{1-\frac{sN(4-\pi)E_s}{2N_0}} \right). 
\end{equation}
The  average SEP of the proposed scheme can be calculated by substituting the above MGF in the SEP expression for $M$-PSK signaling given in \eqref{5}, where we obtain the following for binary signaling $(w_1=0$ and $w_2=\pi) $:
\begin{equation}\label{SEP_2}
P_e= \frac{1}{\pi}  \! \int_{0}^{\pi/2} \! \! \left( \dfrac{1}{1+\frac{N(4-\pi)E_s}{2 \sin^2 \! \eta N_0}}\right) ^{\!\frac{1}{2}}  \! \!\! \exp\left( \dfrac{ -\frac{N^2 \pi E_s}{4  \sin^2 \! \eta N_0}}{1+\frac{N(4-\pi)E_s}{2 \sin^2 \! \eta N_0}} \right) \! \! d\eta. 
\end{equation}
By letting $\eta=\pi/2$ and considering the SNR range of interest 
$\frac{N E_s}{N_0} \ll 10 $,  $P_e$ becomes proportional to
\begin{equation} \label{Approx_3}
P_e \propto \exp \left( -\frac{N^2 \pi E_s}{4 N_0 } \right). 
\end{equation}
Two main results can be inferred from \eqref{Approx_3}. First, the proposed concept in which the LIS acts as an AP, can convey information in an ultra-reliable manner as the dual-hop (DH) LIS scheme given in Fig. 1. Second, by comparing \eqref{Approx_1} and \eqref{Approx_3}, to achieve a target BEP, around $1$ dB improvement in the required SNR can be obtained by the proposed concept compared to the LIS-based DH scheme for binary signaling.

For $M$-ary signaling, substituting \eqref{MGF_2} in  \eqref{5}, we obtain the average SEP in the form of a definite integral as follows:
\begin{align}\label{SEP_3}
P_e= \frac{1}{\pi} & \! \int_{0}^{(M-1)\pi/M} \! \! \left( \dfrac{1}{1+\frac{N(4-\pi) \sin^2(\pi/M)E_s}{2 \sin^2 \! \eta N_0}}\right) ^{\!\frac{1}{2}}  \nonumber 
\\ & \times\exp\left( \dfrac{ -\frac{N^2 \pi  \sin^2(\pi/M) E_s}{4 \sin^2 \! \eta N_0 }}{1+\frac{N(4-\pi)  \sin^2(\pi/M) E_s}{2 \sin^2 \! \eta N_0}} \right) \! \! d\eta. 
\end{align}
Upper bounding this result by letting $\eta= \pi/2$ and focusing on the SNR range of interest, we obtain
\begin{equation} \label{Approx_4}
P_e \propto \exp \left( - \sin^2 (\pi/M) \frac{N^2 \pi E_s}{4 N_0 } \right). 
\end{equation}
Comparing this result with \eqref{Eq:10}, we conclude that a loss  can be expected in the required SNR for higher order signaling $ (M \ge 16) $ due to the SNR loss of $M$-PSK compared to $M$-QAM. However, as will be shown in next section, this loss becomes insignificant considering the potential of the new approach and the relatively low SNR ranges of interest with increasing $N$.

\subsection{Blind Access Point-LIS}
In this worst-case scenario, the LIS does not have the knowledge of channel phases $\psi_i$ and plays the role of a data source by simply adjusting its reflector-induced phase terms in a similar fashion to PSK. To be generalized later, let us focus on the simplest case of binary signaling, in which the reflector-induced phases of the LIS are adjusted for messages 1 and 2 as follows: $\phi_i= \omega_1$ and $ \phi_i= \omega_2 $ for all $i$.
For this scheme, again assuming that the RF source is close enough to the LIS and its transmission is not affected by fading, the received signal in the baseband becomes
\begin{equation}\label{16}
r= \sqrt{E_s}\left[ \sum_{i=1}^{N} g_i \right]  e^{j \omega_m}  + n= \sqrt{E_s} G e^{j \omega_m}  + n
\end{equation}
where $m \in \left\lbrace 1,2\right\rbrace$. For this signal model, following a similar analysis, the CPEP can be obtained as follows:
\begin{equation}
P_{e\left| \right. \! G}= Q\left( \sqrt{\frac{ E_s \left|G \right|^2  \left(1-\cos (\omega_2-\omega_1) \right)}{N_0}} \right). 
\end{equation}
This CPEP expression also requires uniformly distributed phases around the unit circle for the minimization of the SEP. For instance, selecting $w_1=0$ and $w_2=\pi$ as in BPSK will be optimum for binary signaling in terms of BEP.

Noting that $G \sim \mathcal{C N}(0,N)$ under the CLT, the MGF of the instantaneous received SNR $\gamma= \left| G\right|^2 E_s /N_0  $ becomes $ M_{\gamma}(s)=(1-\frac{sNE_s}{N_0})^{-1} $.
Then, substituting $M_{\gamma}(s)$ in \eqref{5} for binary and $M$-ary signaling respectively yields the following average BEP and SEP expressions:
\begin{equation}\label{29}
P_e = \frac{1}{\pi} \int_{0}^{\pi/2} \! \! \left( \! \frac{1}{1+ \frac{NE_s}{ \sin^2 \!\eta  N_0}}  \! \right)  \! d\eta
\end{equation}
and
\begin{equation}\label{30}
P_e=\frac{1}{\pi} \int_{0}^{(M-1)\pi/M} \left( \frac{1}{1+\frac{N \sin^2(\pi/M) E_s}{ \sin^2 \! \eta N_0}}\right)  d\eta.
\end{equation}
It is worth noting that similar to the blind LIS-assisted DH scheme, only an $N$ times SNR gain can be obtained compared to point-to-point transmission over Rayleigh fading channels. This proves that a LIS can be used as an AP as well by only adjusting reflector-induced phases according to the data.

\section{Simulation Results}
In this section, we provide computer simulation results for the LIS-based new (LIS-AP) scheme and make comparisons with the LIS-assisted DH (LIS-DH) scheme. In all simulations, we assume uncorrelated Rayleigh fading channels and consider $E_s/N_0$ as the SNR, similar to the classical diversity combining schemes.

In Fig. 5, we present the BER performance of LIS-DH and LIS-AP schemes for different number of reflecting elements $(N)$ and BPSK signaling along with theoretical curves of \eqref{SEP_2}. As seen from Fig. 5, a LIS can be effectively used as an AP by providing ultra-reliable communications. Furthermore, as verified from \eqref{Approx_3}, around $1$ dB SNR improvement can be obtained compared to the LIS-DH scheme when $M=2$.

In Fig. 6, we evaluate the symbol error rate (SER) performance of LIS-DH and LIS-AP schemes for varying signaling orders $M\!\in \! \left\lbrace 4,16,64 \right\rbrace $ with $64$ reflectors. Theoretical SEP curves obtained from \eqref{M_QAM} and \eqref{SEP_3} are also shown in the same figure to check the accuracy of our theoretical findings. As seen from Fig. 6, both schemes suffer from a degradation in error performance with increasing $ M $, while this is more noticeable for the LIS-AP scheme due to the SNR loss of $M$-PSK over $M$-QAM for $M\ge 16$.

In Fig. 7, we show the BER performance of the blind LIS-DH and LIS-AP schemes for different number of reflectors and BPSK signaling. For comparison, theoretical curves obtained from \eqref{eq:14} are also shown. It is worth noting that both LIS-DH and LIS-AP schemes have the same received SNR distribution for blind transmission, and provide $N$ times SNR gain compared to point-to-point signaling over Rayleigh fading channels. As seen from Fig. 7, doubling $N$ provides a $3$ dB improvement in the required SNR to achieve a target BER value. We also note that although improvements are possible with increasing $N$, the clear advantage of using a LIS diminishes when the intelligence of the surface is not exploited through phase removal in this worst-case transmission scenario.

\begin{figure}[!t]
	\begin{center}
		\includegraphics[width=0.83\columnwidth]{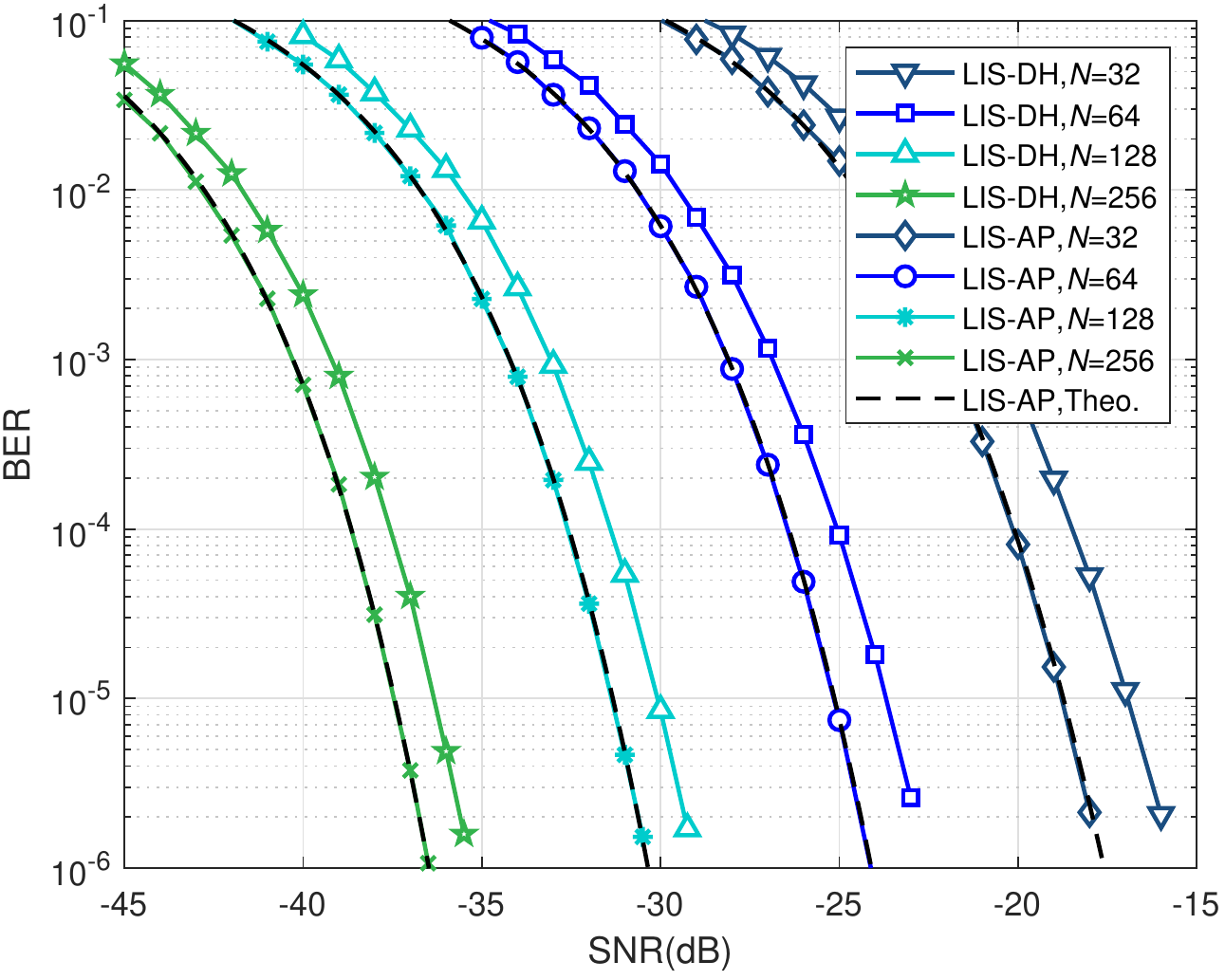}
		\vspace*{-0.35cm}\caption{BER performance of LIS-DH and LIS-AP schemes with varying number of reflectors for BPSK.}\vspace*{-0.4cm}
	\end{center}
\end{figure}

\begin{figure}[!t]
	\begin{center}
		\includegraphics[width=0.83\columnwidth]{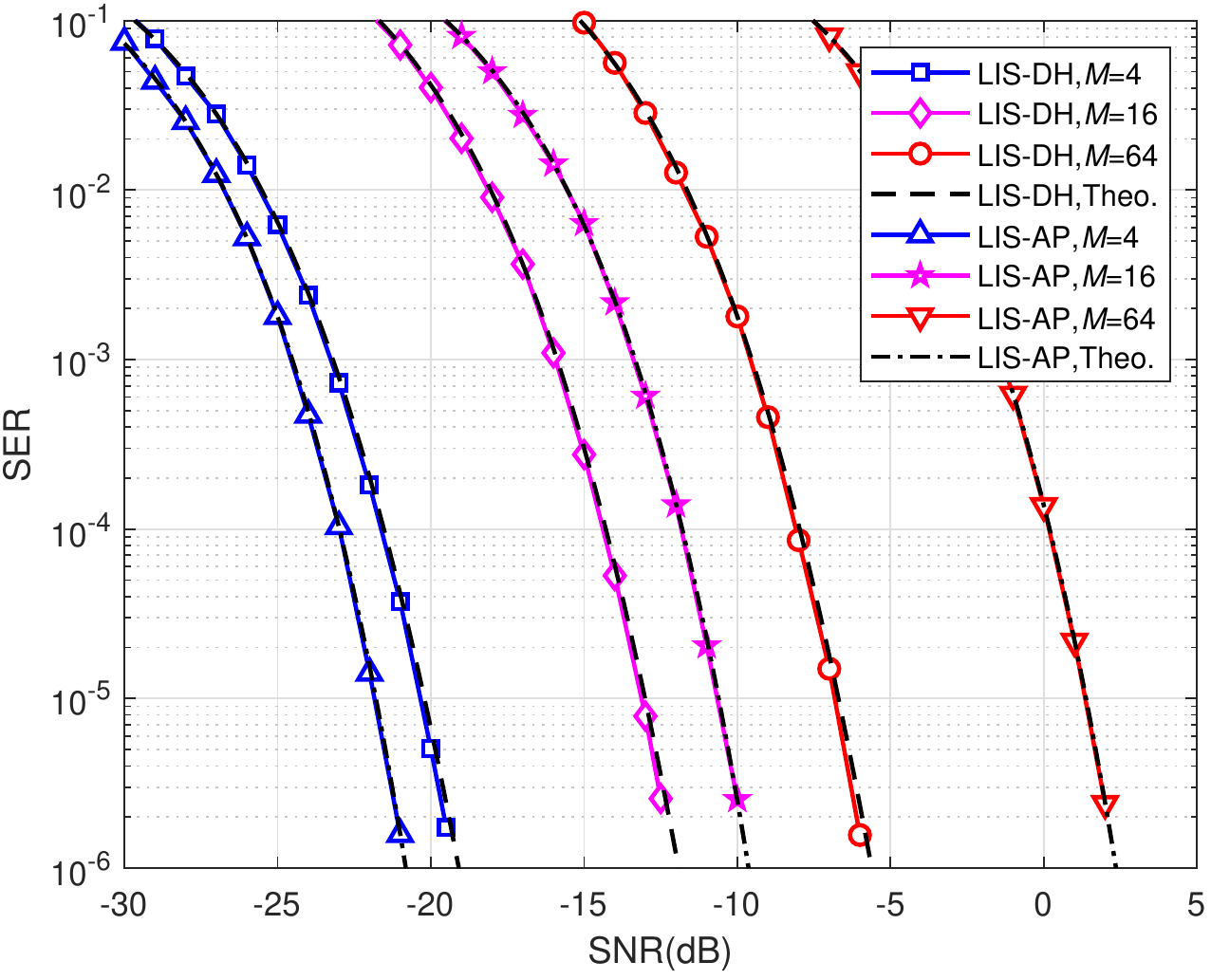}
		\vspace*{-0.35cm}\caption{SER performance of LIS-DH and LIS-AP schemes with $M$-ary signaling for $N=64$.}\vspace*{-0.4cm}
		\label{N_64_SER_ALL}
	\end{center}
\end{figure}

\section{Conclusions}
In this study, we have evaluated the potential of LIS-assisted communications from an error performance perspective and introduced a mathematical framework to assess the potential of this promising approach for future wireless communication systems. Our findings have revealed that LIS-based tranmission can effectively boost the received SNR and enable ultra-reliable communications at extremely low SNR values. We have also shown that the LIS itself can be exploited as an AP with a very simple transceiver architecture, in which an unmodulated carrier is intentionally reflected by the LIS. We conclude that the effective use of a LIS may be a game-changing paradigm in modern communications by eliminating the need for sophisticated coding and massive MIMO schemes. Exploration of multiple-antenna nodes, IM-based schemes, correlated channels, different fading types, and discrete phase shifts appear as interesting future research directions.

\begin{figure}[!t]
	\begin{center}
		\includegraphics[width=0.83\columnwidth]{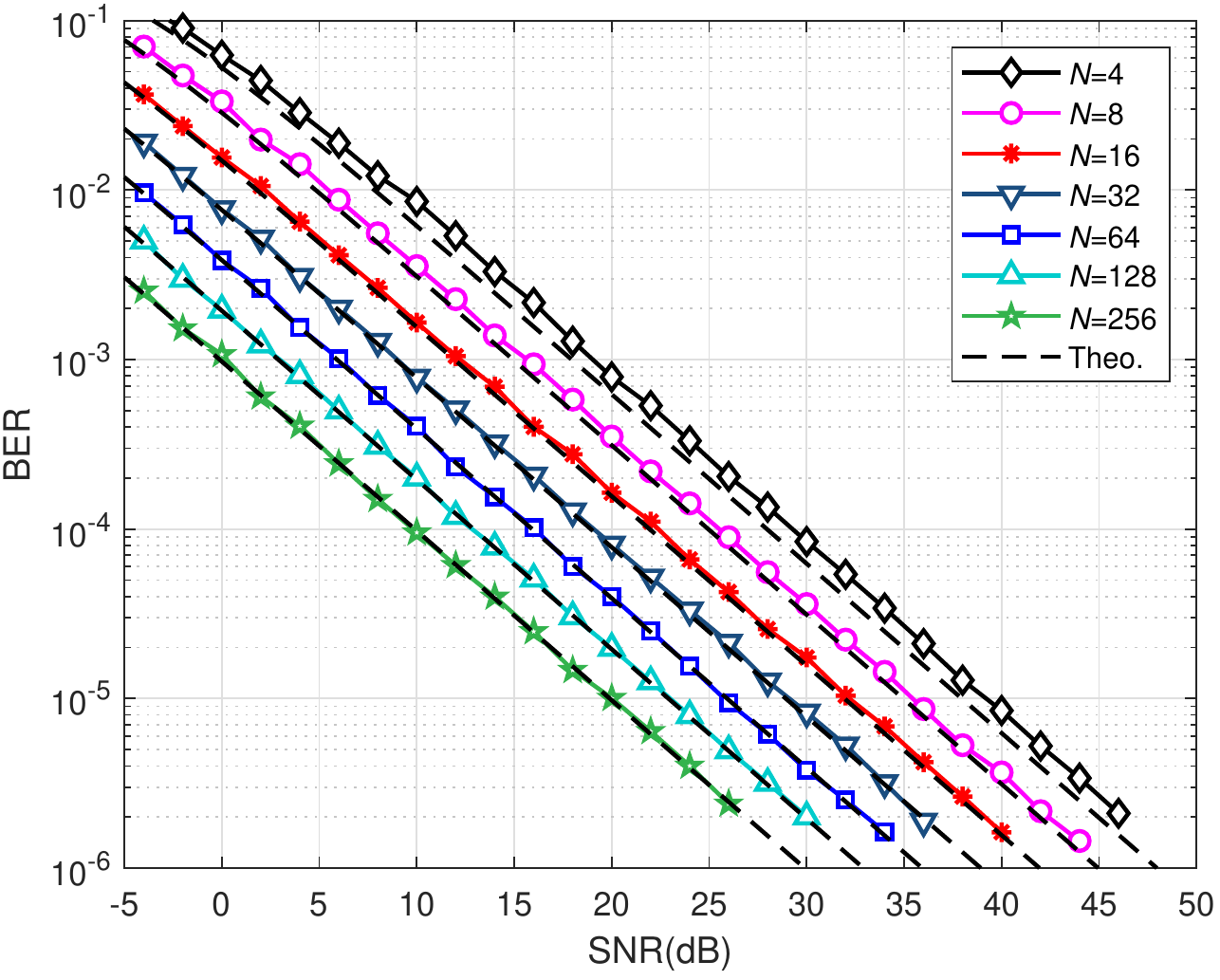}
		\vspace*{-0.35cm}\caption{BER performance of LIS-DH and LIS-AP schemes with blind transmission and varying number of reflectors for BPSK.}\vspace*{-0.4cm}
		\label{N_4_64_BER_BLIND}
	\end{center}
\end{figure}

\bibliographystyle{IEEEtran}
\bibliography{IEEEabrv,bib_2019}

\end{document}